\documentclass[11pt]{revtex4}

\def\al{\alpha}
\def\be{\beta}

\def\de{\delta}
\def\ep{\epsilon}
\def\ze{\zeta}
\def\et{\eta}

\def\ka{\kappa}
\def\la{\lambda}
\def\si{\sigma}
\def\ta{\tau}

\def\Om{\Omega}
\def\mn{{\mu\nu}}
\def\ab{{\al\be}}
\def\du#1{\widetilde{#1}}

\def\fr#1#2{{{#1} \over {#2}}}
\def\frac#1#2{{\textstyle{{#1}\over {#2}}}}
\def\half{{\textstyle{1\over 2}}}

\def\prt{\partial}

\def\inv{^{-1}}

\newcommand{\beq}{\begin{equation}}
\newcommand{\eeq}{\end{equation}}
\newcommand{\bea}{\begin{eqnarray}}
\newcommand{\eea}{\end{eqnarray}}
\newcommand{\rf}[1]{(\ref{#1})}

\def\gT#1#2{{g^{(T)#1}_{#2}}}
\def\gA#1#2{{g^{(A)#1}_{#2}}}
\def\gM#1#2{{g^{(M)#1}_{#2}}}
\def\t{T}

\def\ud#1#2{{{^{#1}_{\phantom{#1}#2}}}}

\def\dugT#1#2{{\du g^{(T)#1}_{#2}}}
\def\dugA#1#2{{\du g^{(A)#1}_{#2}}}

\begin{document}

\title{Finsler-like structures from Lorentz-breaking classical particles}
\begin{abstract}
A method is presented for deducing classical point-particle Lagrange functions 
corresponding to a class of quartic dispersion relations.
Applying this to particles violating Lorentz symmetry 
in the minimal Standard-Model Extension leads to a variety of novel 
lagrangians in flat spacetime.
Morphisms in these classical systems are studied 
that echo invariance under field redefinitions in the quantized theory.
The Lagrange functions found offer new possibilities for 
understanding Lorentz-breaking effects 
by exploring parallels with Finsler-like geometries. 
\end{abstract}

\author{Neil Russell}
\affiliation{Northern Michigan University, Marquette, MI 49855, U.S.A.}
\maketitle

\section{Introduction}
In recent years, 
interest in the possibility of Lorentz violation in nature has blossomed. 
A valuable framework for the study of such 
deviations from the exact predictions of relativity
is the Standard-Model Extension,
or SME,
which provides a systematic accounting of 
hypothesized symmetry-breaking background fields
at the level of the fundamental particles in Minkowski space
\cite{07dcak,08dcak}
and in curved spacetime
\cite{04grav,06Bailey}. 
For a recent review, 
see Ref.\ \cite{14jtReview}.
The focus of this work 
is on the behavior of classical particles 
in spacetime with Lorentz violation.
This classical limit is relevant to the study of 
wave packets, 
macroscopic bodies,
and relativistic scattering,
among other things.

An ambitious goal would be to provide equations for classical systems 
moving under the effects of gravity,
while interacting with each other,
or with an external electromagnetic field,
or with both.
Significant progress towards this goal 
has been made in a general treatment 
of matter-gravity couplings in the presence of Lorentz violation 
\cite{11Tasson},
which includes a detailed study of the dynamics of a test body
in the spin-independent limit of Lorentz violation.
Based on this, 
atomic wave packets in gravitational free fall 
have been used as test bodies 
to investigate the Einstein equivalence principle
with unprecedented precision
\cite{11HohenseeChuEtal},
and 
related work has placed constraints on SME coefficients
by considering atomic systems as classical bound states
\cite{13Hohensee}.
Alternatively, 
intrinsic spin can be highly amplified by  atomic polarization,
providing access to other combinations of Lorentz-breaking matter-gravity couplings.
This has led to innovative experimental tests of Lorentz symmetry 
using the atomic comagnetometer
and the torsion pendulum
\cite{12Tasson,10Romalis,10Panjwani}.
A recent work has explored classical motion 
with spin-dependent isotropic Lorentz violation \cite{14SchreckMinimal}.
Tests of fundamental physics with macroscopic systems in curved spacetime 
with Lorentz violation
pose major technical challenges,
and,
consequently, 
the majority of the limits on coefficients for Lorentz violation 
have been attained in the flat-spacetime limit
\cite{08tables}.
Nevertheless,
searches for novel matter-gravity couplings are well justified 
given that some effects could be accessible only through gravitational couplings 
\cite{09akjtPRL}.

A practical approach to understanding classical behavior 
with Lorentz breaking 
is to consider,
as an intial step,
the flat spacetime limit
in the absence of any potential.
This eliminates complications due to curvature,
removes the complexities of position dependence in the SME coefficients,
and allows existing results from the flat-spacetime SME 
in field theory to be used directly.
The particle trajectory follows from variation of the action $S=\int L d\la$,
where the function $L$ is independent of position $x^\nu$. 
In the resulting equation of motion, 
the velocity $u^\nu$ 
is the derivative of position $x^\nu(\la)$
with respect to the trajectory parameter $\la$.
To ensure that the physics arising from the action is independent of the parametrization, 
the Lagrange function $L$ must be homogeneous of degree one in the velocity $u^\nu$.
The parameter $\la$ can be chosen as the proper time, 
but other choices may be more convenient \cite{10classical}.
The basic goal is then to find the classical Lagrange function $L$
for a particle of mass $m$ propagating at velocity $u^\mu$ in Minkowski spacetime
with constant background coefficients controlling Lorentz-breaking effects.
The validity of the classical limit is established by ensuring 
that the dispersion relation matches that of the fermion in the field theory,
which is known exactly in closed form
\cite{13akmmNonmin}.
For simplicity, 
the present work is restricted to 
the minmal-SME limit, 
for which 
the eight relevant controlling coefficients are 
$a_\nu$, $b_\nu$, $c_\mn$, $d_\mn$, $e_\nu$, $f_\nu$, $g_{\la\mu\nu}$, and $H_\mn$,
and the dispersion relation 
is given by Eq.~(14) of Ref.~\cite{causality}
or, equivalently, by Eq.\ (1) of Ref.~\cite{10classical}.

A method for finding the general 
$L(u;m,a_\nu, b_\nu, \ldots)$ 
exists
\cite{10classical},
and is relevant for minimal and nonminimal Lorentz violation
\cite{14Schreck}.
It involves finding a polynomial equation in $L$ based on the dispersion relation, 
the wave-packet motion, and the homogeneous form of the Lagrange function. 
The multiple roots $L$
reflect the spin up, spin down, matter, and antimatter states of the Dirac system. 
This method
can be challenging to apply in the case of dispersion relations that are 
of higher order than quadratic in the momentum.
For the special case of dispersion relations that are quadratic in the momentum, 
there exists a template that allows the corresponding Lagrange function 
to be deduced immediately \cite{10classical}.

To introduce the ideas and notation, 
a brief review of the quadratic template will be useful.
Dispersion relations with quadratic dependence
on the momentum $p_\nu$
can be put into the form 
\beq
p\Om p + 2\ka\Om p -\mu^2+\ka\Om\ka=0
\, ,
\label{quadraticDR}
\eeq
where 
${\Om^\mu}_\nu$, 
$\ka_\nu$,
and $\mu$ 
are constants that can be viewed as 
a slightly modified identity $\de\ud\mu\nu$, 
a momentum shift, 
and a modified mass. 
Matrix notation is used where possible,
and transposes of single-index objects 
are left to be inferred from the context. 
For example, 
the first term might be written $p^{\rm T}\Om p$ in stricter matrix notation,
or $p_\si {\Om^\si}_\ka p^\ka$using indices.
Raising and lowering is done with the Minkowski metric 
$\et_\mn$ with signature $(+,-,-,-)$.
In the Lorentz-preserving limit, 
${\Om^\mu}_\nu\rightarrow{\de^\mu}_\nu$,
$\ka^\mu\rightarrow 0$,
and 
$\mu\rightarrow m$,
so the relation becomes $p^2=m^2$.
Note that $\Om$ is always invertible, since Lorentz violation is assumed minuscule.
As an example, 
a particle propagating 
in the $a_\nu$ and $f_\nu$
background
has dispersion relation of this
quadratic type,
\beq
(p-a)^2 - m_{af}^2 -(p\cdot f)^2 = 0 
\, . \label{Raf}
\eeq
The mass subscript takes into account 
the possibility that the mass enters differently 
for each Lorentz-breaking background.
The Lagrange function corresponding to 
the dispersion relation 
\rf{quadraticDR} is  
\beq
L(u;\mu,\ka,\Om) = -\mu\sqrt{u\Om\inv u} +\ka\cdot u
\, .
\label{Lquadratic}
\eeq
For the example corresponding to 
Eq.\ \rf{Raf},
$L(u;m_{af},a,f)$
follows with no difficulty.
Ref.\ \cite{10classical}
includes a detailed discussion of this quadratic template
and further examples.
Note, for example, that $L(u,-\mu, \ka,\Om)$ describes the antiparticle after reinterpretation.

One way to lift the Minkowski-space results into the gravitational context
is to employ minimal coupling.
This implies introducing dependence on position $x^\mu$,
thereby promoting the Minkowski metric $\et_\mn$ to 
a pseudo-Riemann metric $g_\mn(x)$,
and converting the SME coefficients 
into functions of position.
The resulting Lagrange function 
$L(x^\mu,u^\mu;m,a_\nu(x),b_\nu(x), \ldots)$
depends on both position and velocity
and describes a classical particle moving under the influence of gravity
and subject to Lorentz breaking. 
The geodesics of the particle follow by
applying the Euler-Lagrange equations.
These ideas form the backdrop to the work presented here.

Numerous parallels exist between 
structures for Lorentz-breaking classical point particles in spacetime
and those of Finsler and pseudo-Finsler geometries.
Some of the broad literature relevant in the present context
includes 
Refs.\ \cite{18Finsler,05bcs,03ShenZermelo,04BRS,06Bogoslovsky,
11pfeifer,11Vacaru,12Laemmerzahl,13mjms,14Silva,11Romero,14Kouretsis}.
The resemblances 
allow results in the existing mathematical literature to be applied in some cases, 
suggest paths to categorizing types of classical Lorentz violation,
and may provide an alternative geometrical framework 
that admits explicit Lorentz violation consistently 
\cite{04grav}.
Note that the signature of the underlying metric 
has a strong impact on the geometry.
In particular, 
the results for Finsler manifolds, 
which have Riemann signature, 
do not necessarily carry over into 
pseudo-Finsler ones.
However,
some ties between the two signatures can 
be made using Wick rotation or other procedures.
Recent work has 
highlighted the link between the Randers space
\cite{41Randers}
and the SME $a_\nu(x)$ background,
identified a calculable Finsler structure 
based on the SME $b_\nu(x)$ background
\cite{11akfinsler},
and found others related to the SME $H_\mn(x)$
coefficient
\cite{12bipartite}. 
Lagrange functions to be presented below 
offer avenues for additional SME-based Finsler 
and pseudo-Finlser structures.

There are two main sets of ideas presented.
The first new result of this work
is a quartic template for finding 
classical Lagrange functions 
$L(u;m,a_\nu,b_\nu,\ldots)$ corresponding to 
a variety of quartic dispersion relations
in flat spacetime.
This extends the existing template 
for the case of quadratic dispersion relations
discussed above,
and is presented in Section \ref{main}.
For dispersion relations that fit the template, 
$L$ can be found 
without the 
contortions of the general method \cite{10classical},
which include  
eliminating momentum variables from a system of equations 
and factoring a high-order polynomial.
Several examples involving single and multiple background fields are discussed
in Section \ref{examples}.

The second set of results 
are mappings 
between classical systems with Lorentz violation.
It is a well-established principle that 
physical results must be  unaltered
under coordinate changes
and field redefinitions.
Consequently, 
some combinations of coefficients in the SME 
are unobservable 
and can be mapped into others.
In the fermion sector,
literature on this topic 
includes Refs.\ 
\cite{07dcak,08dcak,04grav,11Tasson,13akmmNonmin,10Altschul,06Lehnert,02DCPMD}.
A first approach to the question of mappings,
discussed in Section \ref{maps},
is to look at isomorphisms 
between classical systems limited to leading order in Lorentz violation.
Essentially, 
this translates established field-redefinition results 
in the quantum theory 
into corresponding results in the classical approach.
Three particular mappings are relevant, 
between 
the $H$ and $d$ fields, 
between the $b$ field and the axial components of $g$,
and between the $a$ and $e$ fields.
The second approach asks a more challenging question,
whether these three maps can be established between the exact classical systems.
The finding, in section \ref{maps},
is that exact many-to-one correspondences can be set up, 
taking the one field together with a second auxiliary SME field, 
onto the other. 
The choice of auxiliary field is not unique, 
and, in addition, 
mass redefinitions are required.

\section{Classical Lagrangians for quartic dispersion relations}
\label{main}
The new result presented here
is a method for finding classical Lagrange functions 
corresponding to dispersion relations that are quartic in the momentum
and satisfy a particular idempotent condition on the background fields.

Consider a single massive particle in Minkowski spacetime,
with quartic dispersion relation 
\beq
\left[(p+\ka)\Om(p+\ka)-\mu^2\right]^2 = 4 (p+\ka)S(p+\ka)
\, , \label{quarticDR}
\eeq
where 
${S^\mu}_\nu$,
is small, constant, 
and vanishes in the 
Lorentz-preserving limit.
Note that 
the quadratic-template dispersion relation
\rf{quadraticDR}
is recovered by setting $S=0$. 
Limiting attention to dispersion relations 
of this form 
\rf{quarticDR}
with the idempotent condition
\beq
(S\Om^{-1})^2 = \ze (S\Om^{-1})
\, , \label{condition}
\eeq
where $\ze$ is a real constant,
the Lagrange function is
\beq
L(u;\mu,\ka, \Om,S) = -(\mu^2+\ze)^{1/2} \sqrt{u\Om^{-1} u} \mp \sqrt{u\Om^{-1}S\Om^{-1}u} +\ka\cdot u
\, . \label{Lagr}
\eeq
Equations 
\rf{quarticDR},
\rf{condition}, 
and 
\rf{Lagr}
provide a template for finding various Lagrange functions
with the bipartite form \cite{11akfinsler}.
The idea 
is to start from a given quartic dispersion relation
that fits the form \rf{quarticDR},
match coefficients to get the quantities $\Om$, $S$, $\ka$, $\mu^2$,
and arrive at the result \rf{Lagr}
if condition \rf{condition} holds.

As an initial example, 
consider a fermion in an SME background controlled by the $a_\nu$ and $b_\nu$
coefficients.
The dispersion relation can be written
\beq
\left[ (p-a)^2 - (m^2+b^2)\right]^2 = 4 (p-a) \left(bb-b^2 \de\right) (p-a)
\, .
\label{Rab}
\eeq
Matching with Eq.\ \rf{quarticDR}
gives
$\Om=\de$,
${S^\mu}_\nu=b^\mu b_\nu-b^2{\de^\mu}_\nu$,
$\ka_\nu=-a_\nu$,
and 
$\mu^2=m^2+b^2$.
Condition \rf{condition} holds with $\ze=-b^2$,
giving
\beq
L(u;m_{ab},a,b) = -m_{ab}\sqrt{u^2} \mp \sqrt{(b\cdot u)^2- b^2 u^2} - a\cdot u
\, ,
\label{Lab}
\eeq
a result found earlier 
found using other techniques
\cite{10classical}.

Equation \rf{Lagr} can be verified by extracting the canonical momentum
$p_\nu \equiv - \prt L/\prt u^\nu$ from the lagrangian 
$ L\equiv \rho +\si +\ka\cdot u $
and then eliminating the 
velocity $u^\nu$ from these four equations.
This amounts to imposing three conditions,
and the remaining equation constraining the momenta $p_\nu$ 
is the dispersion relation \rf{quarticDR}.
To give some details, 
note that the relationship between the canonical momentum
\beq
p_\nu = -\fr{\mu^2+\ze}\rho (\Om\inv u)_\nu -\fr 1 \si (\Om\inv S\Om\inv u)_\nu -\ka_\nu
\eeq
and the $4$-velocity $u^\nu$
is controlled by the quantities $\Om$, and $S$, and $\ka$. 
Evaluating $(p+\ka)\Om(p+\ka)$ and $(p+\ka)S(p+\ka)$ with the aid of \rf{condition} gives 
\bea
(p+\ka)\Om(p+\ka) &=& 2(\mu^2+\ze)\fr\si\rho +2\ze +\mu^2
\, , \\
(p+\ka)S(p+\ka)&=& \left((\mu^2+\ze)\fr\si\rho +\ze\right)^2
\, .
\eea
The velocity dependence appears as $\si/\rho$ in both
and can be eliminated by substitution to get \rf{quarticDR}.

\section{Examples of classical Lagrange functions}
\label{examples}
\subsection{Case of restricted antisymmetric $d$}
The fermion-sector $d$ coefficient has two indices and is traceless.
Denoting the symmetric part 
by $d_s$, 
the restriction to the antisymmetric components 
is made by setting $d_s=0$.
Using this antisymmetric $d$
to establish conventions, 
the dual is defined by
$\du d_\mn \equiv \half \ep_{\mn\ab}d^\ab$
with 
$\ep_{0123}\equiv-1$.
The standard observer invariants $X$ and $Y$
for antisymmetric tensors
are defined 
\beq
X_d \equiv \frac 1 4 d_{\mn} d^{\mn} 
\, , \ \ \ 
Y_d \equiv \frac 1 4 \du d_{\mn} d^{\mn}
\, ,
\label{XYinvariants}
\eeq
where the subscript identifies the relevant tensor.

The dispersion relation
for the antisymmetric $d$ coefficient
is found to be
\beq
\left[p(\de - d^2)p -m_d^2\right]^2 = 4 m_d^2 pd^2p 
\, ,
\label{Rd}
\eeq
where ${(d^2)^\mu}_\nu = {d^\mu}_\al {d^\al}_\nu$ is obtained by matrix multiplication.

Comparing coefficients with Eq.\ \rf{quarticDR} gives
$
\ka=0
$, $
\Om = \de - d^2
$, $
\mu^2= m_d^2 
$, and $
S = m_d^2 d^2 
$.
The identities
$
d\du d = \du d d = -Y_d \de
$
and 
$
{\du d \mathstrut}^2 = d^2 +2X_d \de
$
can be used to verify the 
antisymmetric-tensor result 
\beq
(\de- d^2)\inv = \fr{\de+{\du d\mathstrut}^2}{1+2X_d-Y_d^2}
\, ,
\label{sqinverse}
\eeq
valid for sufficiently small $d$.
Using this expression,
note that 
$
S \Om\inv = m_d^2 (d^2+Y_d^2 \de)/(1+ 2X_d-Y_d^2)
$
satisfies the projective condition \rf{condition}
if the restriction $Y_d=0$ 
is made.
Evaluation of  $(S\Om\inv)^2$ yields
\beq
\ze = \fr{-2X_d m_d^2}{1+2X_d}
\, .
\eeq
Substituting these results into \rf{Lagr}, 
the classical Lagrange function for the restricted antisymmetric $d$ coefficient is found:
\beq
L(u;m_d,d;d_s=0, Y_d=0) 
= - \fr{m_d}{1+2X_d}\left[\sqrt{u^2+ u {\du d}{\mathstrut\,}^2  u}\pm \sqrt{ud{\mathstrut}^2u}\right]
\, .
\label{Ld}
\eeq 
The Lagrange function for $d$, 
which has a bipartite structure with perturbations occurring in both square roots,
has been studied in Ref.\ \cite{12dcpmd}
using a different approach.

\subsection{Combining $b$ or $H$ with antisymmetric $c$}
The classical Lagrange function $L(u;m_b,b)$ for the $b$ coefficient \cite{11akfinsler},
contained in \rf{Lab},
has bipartite form echoed by 
$L(u;m_H,H;Y_H=0)$ for $H$ with the restriction $Y_H=0$ \cite{10classical}.
Using the quartic template, 
the Lagrange functions that combine the antisymmetric $c_\mn$ coefficient
with each of these can be deduced. 
In these cases, 
the dispersion relation takes the form
\beq
\left[p(\de-c^2)p-m_{cQ}^2+q\right]^2 = 4 p(\de-c)Q (\de+c)p
\, ,
\label{RcQ}
\eeq
where $Q$ is a matrix satisfying the idempotent condition $Q^2=qQ$.
For the $b$ background, 
$Q=bb-b^2\de$ with $q=-b^2$, 
and, for the restricted $H$ background,
$Q=\du H\du H$ with $q=2X_H$.
Equation \rf{RcQ} fits
the quartic dispersion relation template 
\rf{quarticDR}
and matching coefficients yields
$\ka=0$,
$\Om = \de - c^2$,
$\mu^2= m_{cQ}^2-q$,
and
$S = (\de-c)Q(\de+c)$.
Making use of the 
idempotent condition,
it follows that 
$S\Om^{-1}=(\de-c)Q(\de-c)\inv$
satisfies condition \rf{condition}
with 
$\ze=q$.
The result
\rf{Lagr} can be applied, 
leading to 
\bea
&L(u;m_{bc},b,c;c_s=0) = 
-m_{bc}\sqrt{u(\de-c^2){\mathstrut}\inv u} \mp 
\sqrt{u(\de+c){\mathstrut}\inv (bb-b^2\de)(\de-c){\mathstrut}\inv u} \, , & 
\label{Lbc} \\
&L(u;m_{cH},c,H;c_s=0,Y_H=0) = -m_{cH}\sqrt{u(\de-c^2){\mathstrut}\inv u} \mp 
\sqrt{u(\de+c)\inv \du H{\mathstrut}^2(\de-c)\inv u} \, . &
\label{LcH}
\eea
For small $c$, 
the inverses in these expressions exist 
and are found from \rf{sqinverse} 
and
\beq
(\de \pm c)\inv = \fr{\de \mp(c-Y_c \du c)+\du c^2}{1+2X_c-Y_c^2}
\, .
\eeq 
The Lagrange function \rf{Lbc} for $b$ and antisymmetric $c$
describes a classical particle in a Lorentz-breaking background 
with ten independent degrees of freedom: four from the components of $b_\nu$
and six from the antisymmetric components of $c_\mn$.
In the case of \rf{LcH}, there are eleven independent components breaking Lorentz symmetry: 
six from each of the two antisymmetric tensors, 
with a reduction of one due to the $Y_H=0$ condition.

\subsection{Axial and Trace components of the $g$ coefficient}
Using notation from Ref.\ \cite{12fittante}, 
the $g_{\mu\nu\la}$ coefficient has a unique decomposition 
$g_{\mu\nu\la}  = \gA{}{\mu\nu\la}+\gT{}{\mu\nu\la}+\gM{}{\mu\nu\la}$
where the axial part $\gA{}{\mu\nu\la}$ and the 
trace part $\gT{}{\mu\nu\la}$ 
have $4$ components each, 
and the mixed-symmetry part $\gM{}{\mu\nu\la}$ has $16$ components.

The dispersion relation for the $g$ coefficient
is \cite{causality}
\beq
0=(p^2+m_g^2-2X_{gp})^2 - 4 m_g^2 p^2 - 4 p(gp)^2p + 4 Y_{gp}^2
\, ,
\label{Rg}
\eeq
where 
$(gp)_\mn \equiv g_{\mu\nu\la}p^\la \equiv g_\mn(p)$,
and the definitions of $X$ and $Y$ follow the convention 
in Eq.\ \rf{XYinvariants},
$X_{gp}=g_\mn(p)g^\mn(p)/4$,
and
$Y_{gp}=g_\mn(p)\widetilde g^\mn(p)/4$.

Applying the constraint $\gM{}{\mu\nu\la}=0$, 
attention is limited to cases where only the axial and trace components 
are nontrivial.
These can be expressed in terms of $4$-vectors,
\beq
A^\al \equiv \gA\al{} \equiv \frac 1 6  g_{\si\ka\ta}\ep^{\si\ka\ta\al}
\, , \qquad
\t_\nu \equiv \frac 1 3 \gT{}\nu \equiv \frac 1 3 {g_{\nu\al}}^\al
\, ,
\label{vecdefs}
\eeq
and, 
together with the following 
tensor containing information about the momentum,
\beq
{P^\mu}_\nu \equiv p^\mu p_\nu - p^2 \de^\mu_\nu
\, , \label{Pdef}
\eeq
can be used to simplify the dispersion relation
\rf{Rg}.
It is found that 
$
p(gp)^2p = p^2 \t P \t
$
as a result of the total antisymmetry of 
$\gA{}{\mn\al}$.
To evaluate $X_{gp}$ and $Y_{gp}$, 
first note that 
\beq
\gT{\mn}{}(p) = \t^\mu p^\nu - \t^\nu p^\mu 
\, , \qquad
\gA{}{\mn}(p) = \ep_{\mn\al\be} A^\al p^\be
\, .
\eeq
Substitution into the expression for $X_{gp}$
is aided by the identities
\bea
\gA{}\mn(p) \gT\mn{}(p) &=& 0 
\, , \nonumber\\
\half\gA{}\mn(p) \gA\mn{}(p) &=& (A\cdot p)^2 - A^2 p^2 = APA
\, , \nonumber\\
\half\gT{}\mn(p) \gT\mn{}(p) &=& -(\t\cdot p)^2 + \t^2 p^2 = - \t P\t
\, , \label{Xgidents}
\eea
and substitution into the expression for $Y_{gp}$
is facilitated by the identities
\bea
\half\gA{}\mn(p) \dugT\mn{}(p) &=& (\t\cdot p)(A\cdot p)-(\t\cdot A)p^2 = \t P A
\, , \nonumber\\
\gA{}\mn(p) \dugA\mn{}(p) &=& 0
\, , \nonumber\\
\gT{}\mn(p) \dugT\mn{}(p) &=& 0
\, , \label{Ygidents}
\eea
giving
\beq
X_{gp} = \frac 1 2 APA - \frac  12 \t P \t
\, , \quad
Y_{gp} = \t P A 
\, .
\eeq
The dispersion relation \rf{Rg} for the axial and trace components of $g$
takes the form
\beq
0 = (p^2 - m_{AT}^2 - APA - \t P \t)^2 - 4 m_{AT}^2 APA + 4(\t PA)^2 - 4 (APA)(\t P \t)
\, .
\label{Rg2}
\eeq
The last two terms are quartic in the momentum and vanish if either $A^\mu=0$ or $\t^\mu=0$.
In these limits, the form matches the template \rf{quarticDR}.

To find the Lagrange function for the special case where only the trace components of $g$ 
are nonzero,  
the axial vector $A^\mu$ in \rf{Rg2} is set to zero yielding the quadratic expression 
$p\Om p=\mu^2$,
where 
${\Om^\mu}_\nu=\de^\mu_\nu - \t^\mu \t_\nu/(1+\t^2)$
with inverse 
${{(\Om\inv)}^\mu}_\nu=\de^\mu_\nu+\t^\mu\t_\nu$,
and 
$\mu=m_T/\sqrt{1+\t^2}$.
The Lagrange function is found using the template \rf{Lquadratic}
with $\ka=0$:
\beq
L(u;m_T,T) = \fr{-m_T}{\sqrt{1+\t^2}}\sqrt{u^2+(\t\cdot u)^2}
\, .
\label{LgT}
\eeq
Note that 
the Finsler structure discussed in Ref.\ \cite{89Beil}
has a similar structure if restricted to a single tangent space.

The Lagrange function for the $g$ coefficient with only axial components
can be found by setting $\t^\mu=0$ in \rf{Rg2}.
A match with the template dispersion relation \rf{quarticDR}
can be achieved with 
$\ka=0$,
$\Om=(1+A^2)\de-AA$,
$S=m^2(AA-A^2\de)$,
and 
$\mu^2=m_A^2$.
For small $A^\mu$, 
the inverse 
${\Om\inv} =(\de+AA)/(1+A^2)$
exists.
The idempotent property \rf{condition}
can be confirmed after evaluating 
$S\Om\inv$, 
and it is found that 
$\ze=-m_A^2 A^2/(1+A^2)$.
The resulting Lagrange function for axial $g$ is
\beq
L(u;m_A,A) = -\fr {m_A}{1+A^2}\left[\sqrt{u^2+(A\cdot u)^2} \pm \sqrt{(A\cdot u)^2- A^2 u^2}\right]
\, .
\label{LgA}
\eeq

As with Eq.\ \rf{Ld},
this Lagrange function has a bipartite form
with SME coefficients in both the square roots.
It is interesting to compare it
with the Lagrange function 
$L(u;m_b,b)$
contained in \rf{Lab}.
In the quantum context, 
a field redefinition relates $b^\mu$ to $A^\mu$
at leading order. 
In the next section,
this relationship is investigated in the classical-particle context.
Also note that the couplings of 
$b^\mu$ or $A^\mu$ coefficients to fermions
have the same form as 
matter-torsion couplings 
in Riemann-Cartan gravity,
and this has allowed results from Lorentz tests 
to place tight constraints on 
spacetime torsion
\cite{08torsion,13NeutronSpin}.

The Lagrange functions presented in this section 
involve uniform Lorentz-breaking background fields 
in Minkowski spacetime.
In this limit, classical particles undergo no acceleration, 
because of the homogeneity of the Lagrange function
and the absence of position or parameter dependence \cite{10classical}.
This means constant Lorentz violation is unobservable with a single point particle.
However, comparison of systems with distinct properties 
makes physical effects measurable.
In curved spacetime, 
where the coefficients for Lorentz violation depend on position and time,
the Lagrange functions can be viewed as Finsler-like structures. 
The corresponding particles follow geodesics
controlled by the background fields and the curvature of the space.
Geodesic equations for the $b_{\mu}(x)$ and $H_{\mn}(x)$ backgrounds
are known \cite{11akfinsler,12bipartite},
and those for others such as the ones presented in this section 
are left for future work. 
The important feature of the present Lagrange functions 
is that they correspond to known SME dispersion relations,
so their further study in the Finsler context is of definite interest.

\section{Field redefinitions and maps}
\label{maps}
In the SME, 
the freedom to redefine fields
implies that some Lorentz-breaking fields 
can be mapped into others. 
The focus here is on the three mappings 
that pair the dimension-$3$ operators 
in the minimal SME
with dimension-$4$ operators 
at leading order
in the quantized theory.
At first order in Lorentz violation, 
they are implemented by the replacements 
\bea
\du H_\mn &\leftrightarrow& m d_\mn 
\, , \label{Htod}\\
b_\nu &\leftrightarrow& - m A_\nu
\, , \label{btoA}\\
a_\nu&\leftrightarrow& - m e_\nu
\, . \label{atoe}
\eea
In this section, each of these correspondences 
is verified 
in the case of the relevant dispersion relations 
and classical Lagrange functions at leading order in Lorentz violation.
A second, more challenging, issue is also addressed in each case, 
that of whether mappings between the classical systems
that hold exactly at all orders in Lorentz violation
are possible.
The finding is that this can be done
as a many-to-one map
by introducing an appropriate auxiliary background field
in each case.

\subsection{Relating $H_{\mu\nu}$ to antisymmetric $d_{\mu\nu}$}
First, 
consider mappings between the dispersion relations 
at leading order in the Lorentz-breaking $H_\mn$ and $d_\mn$ fields.
The $H$ dispersion relation for $Y_H=0$ can be found from Eq.\ \rf{RcQ}
with appropriate substitutions for $Q$ and $q$,
\beq
\left(p^2-m_{H}^2+2X_H\right)^2 = 4 p{\du H}^2 p
\, ,
\label{RH}
\eeq
while the $d$ dispersion relation for $Y_d=0$  is given in Eq.\ \rf{Rd}.
To express \rf{RH} at first order in Lorentz violation, 
take the square root of both sides, 
and cross out the second-order $2X_H$ term.
The resulting expression appears in Table \ref{mappingtable} 
in the first entry of the second column.
There is no subscript on $m$ 
because a mass correction can only enter at second order.
Immediately below this in the table
is the first-order limit of 
the dispersion relation \rf{Rd} for $d$,
which is found in a similar way.
It follows by inspection that \rf{Htod},
which also appears in the initial column of the table,
relates these first-order 
dispersion relations.

Next, 
consider mappings between the corresponding Lagrange functions
at first order in Lorentz violation.
Noting that $\sqrt{u^2+u{\du d}{\mathstrut}^2u}\approx\sqrt{u^2}$
to first order in $d$,
Eq.\ \rf{Ld} becomes
\beq
L(u;m,d;d_s=0, Y_d=0) 
\approx - m\left(\sqrt{u^2\mathstrut}\pm \sqrt{ud{\mathstrut}^2u}\right)
\, .
\eeq
The Lagrange function for $H$ 
follows from Eq.\ \rf{LcH} with $c=0$,
\beq
L(u;m,H;Y_H=0) = -m\sqrt{u^2\mathstrut} \mp \sqrt{u\du H{\mathstrut}^2 u}
\, ,
\eeq
and it follows by inspection that 
\rf{Htod} relates these leading-order classical Lagrange functions,
which appear in the first row and last column
of Table \ref{mappingtable}.

\begin{table}
\begin{tabular}{|@{\hspace{4pt}}c@{\hspace{4pt}}|@{\hspace{4pt}}c@{\hspace{4pt}}|@{\hspace{4pt}}c|}	
\hline										
{\bf Field map}&{\bf First-order dispersion relations}&\multicolumn{1}{c|}{{\bf First-order Lagrange functions}} \\
\hline
$\du H_\mn \leftrightarrow m d_\mn $&$p^2-m^2 \approx \pm 2\sqrt{p{\du H}{\mathstrut}^2 p}	$&$L(u;m,H;Y_H=0) = -m\sqrt{u{\mathstrut}^2} \mp \sqrt{u\du H{\mathstrut}^2 u}$\\
&$p^2-m^2\approx \pm 2 m \sqrt{pd{\mathstrut}^2 p}$&$L(u;m,d;d_s=0, Y_d=0)\approx - m\left(\sqrt{u{\mathstrut}^2}\pm \sqrt{ud{\mathstrut}^2u}\right)$\\
\hline
$b_\nu\leftrightarrow - m A_\nu$&$p^2-m^2 \approx \pm 2\sqrt{(b\cdot p){\mathstrut}^2- b^2 p^2}$&$L(u;m,b) = -m\sqrt{u^2} \mp \sqrt{(b\cdot u){\mathstrut}^2- b^2 u^2}$\\
&$p^2-m^2 \approx \pm 2 m\sqrt{(p\cdot A){\mathstrut}^2- A^2 p^2}$&$L(u;m,A) \approx - m \left(\sqrt{u^2} \pm \sqrt{(A\cdot u){\mathstrut}^2- A^2 u^2}\right)$\\
\hline
$a_\nu\leftrightarrow - m e_\nu$&$p^2-m^2 \approx 2 a\cdot p$&$L(u;m,a) = -m\sqrt{u^2} -a\cdot u$\\
&$ p^2-m^2\approx -2 m e\cdot p$&$L(u;m,e) \approx -m\sqrt{u^2} + m e\cdot u$\\
\hline
\end{tabular}
\caption{\label{mappingtable}Mappings between dispersion relations and between classical Lagrange functions 
at first order in Lorentz-breaking background fields.}
\end{table}

A natural progression 
is to seek a mapping 
between the full dispersion relations \rf{RH} and \rf{Rd}, 
without taking the leading-order limits.
The replacement \rf{Htod} 
is insufficient to do this.
However,  
if the $H$ dispersion relation 
is augmented to include the antisymmetric $c$ coefficient
as in \rf{RcQ} with $Q=\du H^2$, 
an exact map to the $d$ dispersion relation \rf{Rd} 
is indeed possible. 
It is implemented by the replacements
\bea
\du H_\mn &\rightarrow& \fr{m_d}{\sqrt{1+2X_d}}  d_\mn 
\, , \nonumber\\
c_\mn &\rightarrow& d_\mn 
\, , \nonumber\\
m_{cH} &\rightarrow& \fr{m_d}{\sqrt{1+2X_d}} 
\, .
\label{cH_to_d}
\eea
Note that the correspondence implies 
$
X_H \rightarrow - m_d^2 X_d /(1+2X_d)
$.
The identity $\du{\du H}=-H$ 
means the first expression in \rf{cH_to_d} can be written
$
H_\mn\rightarrow - m_d \du d_\mn/\sqrt{1+2X_d}
$.

These replacements also implement an exact mapping from 
the classical Lagrange function 
$L(u;m_{cH},c,H;c_s=0, Y_H=0)$ 
in Eq.\ \rf{LcH} 
to the Lagrange function 
$L(u;m_d,d;d_s=0,Y_d=0)$
in Eq.\ \rf{Ld}.

This result shows that a mass redefinition 
is part of the morphism relating the different 
Lorentz-breaking systems.
Although $Y_c$ does not need to vanish, 
it can be imposed 
without affecting the replacement $c_\mn\rightarrow d_\mn$,
thereby reducing by one 
the number of independent auxiliary variables needed.
With this assumption,
$H_\mn$, 
$c_\mn$, and 
$d_\mn$ 
each have five independent variables,
and the map \rf{cH_to_d} can be viewed as a projection from the 
$11$-dimensional space 
$(m_{cH},H_\mn,c_\mn)$ 
to the $6$-dimensional space 
$(m_d,d_\mn)$. 
Since the mapping is many-to-one, 
it is not invertible.

\subsection{Relating $b_\nu$ to $g^{(A)}_\nu$ }
To investigate the classical version of the mapping \rf{btoA}, 
the first-order dispersion relations for $b$ and $A$
are needed.
The expression for $b$ 
is found by setting $a=0$ in \rf{Rab} 
and keeping only the first order terms in $b$.
The result appears in 
Table \ref{mappingtable}, row two, column two.
Immediately below it
is the dispersion relation for the $A$ background, 
obtained from the exact result
\rf{Rg2} with $T=0$,
\beq
0 = (p^2 - m_A^2 - APA)^2 - 4 m_A^2 APA
\, ,
\label{RgA}
\eeq
by taking the first-order limit.
Inspection of the entries in the table shows that 
\rf{btoA}
maps the two leading-order dispersion relations.

To verify the mapping \rf{btoA} at the level of the classical Lagrange functions, 
note the following.
The Lagrange function \rf{Lab} with $a=0$ involves 
a correction to the conventional Lorentz-preserving Lagrange function 
that is already first order in Lorentz violation $b$,
and appears in 
Table \ref{mappingtable},
row two, column three. 
The first-order approximation to the $A_\nu$ Lagrange function  \rf{LgA},
\beq
L(u;m,A) \approx - m \left(\sqrt{u^2} \pm \sqrt{(A\cdot u)^2- A^2 u^2}\right)
\, ,
\eeq
appears below it in the table.
By inspection of the table entries, 
it can readily be seen that the replacement \rf{btoA} 
establishes an isomorphism between the $b_\nu$ and $A_\nu$ Lagrange functions 
at first order.

With the use of an additional SME background field, 
$f_\mu$,
an exact mapping can be established
between the all-orders dispersion relation 
for $b$ and $f$, 
\beq
0 = \left(p^2 - (f\cdot p)^2 - m_{bf}^2- b^2\right)^2 - 4 b P b
\, ,
\label{Rbf}
\eeq
and that of $A$, 
given in Eq.\ \rf{RgA}.
The map 
is defined by the replacements
\bea
b_\nu&\rightarrow& \fr {m_A A_\nu}{1+A^2}
\, , \nonumber\\
f_\nu&\rightarrow& \fr{A_\nu}{\sqrt{1+A^2}}
\, , \nonumber \\
m_{bf} &\rightarrow& \fr{m_A}{1+A^2}  
\, .
\label{bf_to_A}
\eea

As with the morphism in \rf{cH_to_d},
this map 
involves a mass redefinition
and an auxiliary field.
Note that the auxiliary $f_\nu$ 
has the same number of independent components 
as $b_\nu$ and $A_\nu$.
This property is also true for the $c_\mn$
auxiliary field used in mapping from $H_\mn$ to $d_\mn$.
The morphism \rf{bf_to_A}
is many-to-one, 
taking the $9$-dimensional space $(m_{bf}, b_\nu, f_\nu)$
to the $5$-dimensional space $(m_A,A_\nu)$.

\subsection{Relating $a_\nu$ to $e_\nu$ }
The dispersion relations for 
a particle in the 
$a$ background or in the $e$ background 
are
\bea
(p-a)^2 - m_a^2 &=& 0 \, , \label{Ra}
\\
p^2 - (m_e -e\cdot p)^2 &=& 0 
\, . \label{Re}
\eea
The first-order approximations 
given in row three, column two
of Table \ref{mappingtable}, 
follow by 
dropping the subscripts on the masses
and keeping only terms that are linear in Lorentz violation.
It follows by inspection of these entries in the table that the replacement
\rf{atoe} implements the map between the two 
leading-order dispersion relations.

Since the dispersion relations \rf{Ra} and \rf{Re} 
are quadratic in the momentum, 
the corresponding classical Lagrange functions follow 
by application of the quadratic template, and are known to be
\bea
L(u;m_a,a) &=& -m_a\sqrt{u^2} -a\cdot u
\, , \label{La} \\
L(u;m_e,e) &=& -\fr{m_e}{1-e^2}\left\{\sqrt{u^2   +(e\cdot u)^2 - e^2 u^2} - e\cdot u \right\}
\, . \label{Le}
\eea
Table \ref{mappingtable}
contains $L(u;m,a)$ in row three, column three 
since it is already first-order in Lorentz violation.
Immediately below it 
is the first-order approximation to $L(u;m,e)$.
Inspection of these entries shows that 
the replacement \rf{atoe}
maps the leading-order Lagrange functions into each other,
as it does in the quantum case.
 
Next, 
consider the question of finding an exact mapping
between the $a$ and $e$ systems,
linking the dispersion relations 
\rf{Ra} and \rf{Re},
and the Lagrange functions 
\rf{La} and \rf{Le}.
As seen in the earlier examples involving the maps 
\rf{cH_to_d} and \rf{bf_to_A},
this can be done with the assistance of an auxiliary field.
The auxiliary field is not unique, 
and to confirm this, 
exact mappings are given using two different auxiliary fields. 

A first mapping to 
the full $e$ dispersion relation 
\rf{Re}
may be established by using 
$\t_\nu$
as the auxiliary field.
To find the dispersion relation for the combined $a$ and $\t$
coefficients, 
the steps leading to \rf{Rg2} can be repeated with a few modifications, including 
the assumption 
$A_\nu=0$.
The result is 
\beq
0 = ((p-a)^2-m_{aT}^2-\t P \t)^2+ 4 pMp
\, ,
\label{RaT}
\eeq
where 
$M\ud\mu\nu $ is defined by
$M = ((a\cdot\t)^2- a^2 \t^2)\de + a^2 \t\t - (a\cdot\t)(a\t+\t a)+ \t^2 a a
$. 
An exact map 
from this dispersion relation for $a$ and $\t$
to that for $e$,
Eq.\ \rf{Re}, is found to be
\bea
a^\mu &\rightarrow& \fr{-m_e e^\mu}{1-e^2} \, , \nonumber\\
\t^\mu &\rightarrow& \fr{e^\mu}{\sqrt{1-e^2}} \, , \nonumber\\
m_{aT} &\rightarrow& \fr{m_e}{1-e^2} \, . 
\label{aT_e}
\eea

To verify that this map works at the level of the classical Lagrange functions, 
$L(u;m_{aT},a,T)$ is needed.
While the dispersion relation \rf{RaT}
is quartic in the momentum, 
it does not appear to fit the quartic template in Sec.\ \ref{main},
and this Lagrange function is unknown at present.

Having seen that $\t$ can serve as an auxiliary field to 
implement an exact map from the $a$ system to the $e$ system, 
the next point to be made is that this auxiliary is not unique. 
To demonstrate this, 
the $f_\nu$ field is used as an alternative auxiliary
to implement the $a$ to $e$ mapping.
The dispersion relation for the combined $a$ and 
$f$ coefficients
is given in Eq.\ \rf{Raf},
and an exact mapping to \rf{Re} 
is made using the replacements 
\bea
a_\nu &\rightarrow& - m_e e_\nu \, , \nonumber\\
f_\nu &\rightarrow& e_\nu \, , \nonumber\\
m_{af} &\rightarrow& \sqrt{1+e^2} \, m_e \, .
\label{af_e}
\eea

To show that these replacements also map
the Lagrange functions, 
$L(u;m_{af},a,f)$ is needed.
Since the corresponding dispersion relation \rf{Raf}
is quadratic, 
the template \rf{Lquadratic} can be applied and the result is
\beq
L(u;m_{af},a,f) = -\sqrt{m_{af}^2 + \fr{(a\cdot f)^2}{1-f^2}} \sqrt{u^2 + \fr{(f\cdot u)^2}{1-f^2}}
-a\cdot u - \fr{a\cdot f}{1-f^2}f\cdot u  
\, . \label{Laf}
\eeq
It can be verified
that the substitutions \rf{af_e} implement 
a map from \rf{Laf} to \rf{Le}.

Note the properties seen in the 
\rf{cH_to_d} and \rf{bf_to_A} morphisms
are echoed in both cases here.
Exact mappings from $a_\nu$ to $e_\nu$
can be constructed
with the aid of an auxiliary field
with the same number of components as 
$a$ and $e$.
In \rf{aT_e} this is the $\t_\nu$ field, 
and in \rf{af_e} it is the $f_\nu$ field,
demonstrating that the auxiliary is not unique.
The map also involves a mass redefinition.
It is many-to-one,
and takes $9$ variables consisting of a mass, 
the four components $a_\nu$, 
and the four components of the auxiliary field, 
to the $5$ variables $(m_e,e_\nu)$.

\section{Summary and Discussion}
The main result presented in section \ref{main}
is a classical Lagrange function in Minkowski space 
that describes a massive particle in the context of minimal Lorentz violation 
with certain quartic dispersion relations.
The result is given in the form of a template:
any particle with dispersion relation matching \rf{quarticDR},
and with Lorentz-breaking background satisfying the idempotent condition 
\rf{condition},
has Lagrange function given in equation \rf{Lagr}.

The quartic template is used to provide explicit Lagrange functions 
for several minimal-SME Lorentz-breaking backgrounds that have not appeared in the literature to date. 
The results for the functions 
$L(u;m_{bc},b,c,c_s=0;c_s=0)$,
and 
$L(u;m_{cH},c,H;c_s=0,Y_H=0)$,
show that Lorentz violation matching the quartic template 
and combining more than one background field 
can be studied with relative ease.
Earlier work \cite{10classical}
provided a template for Lagrange functions with quadratic dispersion relations
and included the explicit form of $L(u;m_{aef},a,e,f)$, 
which has twelve independent components for the background fields.
The quartic examples in this work provide explicit Lagrange functions 
in \rf{Lbc} and \rf{LcH} with ten and eleven independent background-field components respectively.
The Lagrange function $L(u;m_T,T)$ involving the four trace components
of the SME $g$ background,
denoted $T_\nu$,
is given in equation  \rf{LgT}.
In the case of Lorentz violation by the axial components $A_\nu$ of the $g$
background, 
$L(u;m_A,A)$ appears in equation \rf{LgA}.
Finding the classical Lagrange function for all the 24 independent components 
of the $g$ background field remains an open challenge.

In section \ref{maps},
mappings between SME background fields have been studied. 
The leading-order isomorphisms 
between $\du H$ and $d$, 
between $b$ and axial $g$, 
and between $a$ and $e$
are known in the fundamental field-theoretic formalism
to arise because of the freedom to redefine fields and coordinates.
The same leading-order correspondences  are confirmed in the classical limit, 
and summarized in Table \ref{mappingtable}.

The second aspect studied in section \ref{maps}
is exact mappings between the above minimal-SME background fields.
Isomorphisms are not expected, because 
each SME field at the level of the fundamental Lagrange function 
controls a different field operator. 
It is found that exact projective mappings do exist, 
in each case taking more than one SME field onto another SME field.
For the three leading-order mappings already discussed, 
exact projective mappings are established with the aid of 
a second SME background field, 
which at least doubles the number of independent field components
in the domain space.
A feature of these projective maps is 
a rescaling of the mass that depends on the fields. 
This means the mass can be considered an independent component on the same footing 
as the field components.
For the $H$ to $d$ map, 
the antisymmetric components of the $c$ background 
provide an auxiliary field, so that the domain $(m_{Hc},H_\mn,c_\mn)$ with $Y_H=0$ has dimension 12
and maps to the range $(m_d,d_\mn)$ with $Y_d=0$, of dimension 6.
For the $b_\nu$ to $A_\nu$ map, 
the field $f_\nu$ can be used as an auxiliary
and the projection takes the $9$-dimensional domain $(m_{bf},b_\nu,f_\nu)$ onto the 
$5$-dimensional range $(m_A,A_\nu)$.
The auxiliary field used in these projective mappings 
is not unique, 
as can be seen in the case of the 
$a$ to $e$ map, 
for which $f_\nu$ is one possible auxiliary
and $T_\nu$, the trace component of $g$, is another. 
This gives a projection from the $9$-dimensional domain space $(m_{af},a_\nu,f_\nu)$ or $(m_{aT},a_\nu,T_\nu)$
onto the $5$-dimensional range space $(m_e,e_\nu)$. 

The quartic template 
in Eqs.\ \rf{quarticDR}, \rf{condition}, and \rf{Lagr}
is applicable in a limited set of cases
and a natural question is how to generalize it. 
Staying within the minimal SME, 
one direction for further investigation would be to seek ways to weaken 
the idempotent condition \rf{condition}.
This would be useful, 
for example, 
in addressing the Lagrange functions $L(u;m_{bf},b,f)$
and $L(u;m_{aT},a,T)$,
where the respective dispersion relations, 
\rf{Rbf} and \rf{RaT}, match \rf{quarticDR}, 
but the background fields do not satisfy the idempotent condition \rf{condition}.
Another goal in the minimal SME context 
would be to broaden the template to more general quadratic 
dispersion relations, covering for example equation \rf{Rg2}, the case of $g_{\la\mn}$ 
limited to axial and trace components.
The most ambitious goal in the minimal SME would be to find the Lagrange function 
corresponding to the full dispersion relation appearing in Ref. \cite{causality}. 
This is likely to be a very long and unenlightening expression, 
given that even the case of $L(u;m_H,H)$ for a general $H$ background with nonzero $Y_H$ 
is known to be highly complex \cite{10classical}.
There are numerous options for further development. 
Other investigations might include 
methods for finding Lagrange functions for nonminimal Lorentz violation \cite{13akmmNonmin}, 
or for interacting particles.

The main motivation of this work is to provide new 
Finsler-like structures
that are rooted in the Lorentz-breaking background fields of the minimal SME.
The new Minkowski-space classical Lagrange functions found here 
can be used as the basis for Finsler and pseudo-Finsler structures in future work.
They hold the promise of providing insights into the geometrical nature of Lorentz violation 
in classical systems and of gaining new insights into the physical content based on 
existing approaches in Finsler geometry. 

\acknowledgments
Thanks to the Indiana University Center for Spacetime Symmetries for hospitality 
while this research was undertaken.


\begin{thebibliography}{xx}

\bibitem{07dcak}
D.\ Colladay and V.A.\ Kosteleck\'y,
Phys.\ Rev.\ D {\bf 55}, 6760 (1997)
[hep-ph/9703464].

\bibitem{08dcak}
D.\ Colladay and V.A.\ Kosteleck\'y,
Phys.\ Rev.\ D {\bf 58}, 116002 (1998)
[hep-ph/9809521].

\bibitem{04grav}
V.A.\ Kosteleck\'y, 
Phys.\ Rev.\ D {\bf 69}, 105009 (2004)
[hep-th/0312310].

\bibitem{06Bailey}
Q.G.\ Bailey and V.A.\ Kosteleck\'y, 
Phys.\ Rev.\ D {\bf 74}, 045001 (2006)
[gr-qc/0603030].

\bibitem{14jtReview}
J.D.\ Tasson, 
Rep.\ Prog.\ Phys. {\bf 77}, 062901 (2014)
[arXiv:1403.7785].

\bibitem{11Tasson}
V.A.\ Kosteleck\'y and J.D.\ Tasson, 
Phys.\ Rev.\ D {\bf 83}, 016013 (2011)
[arXiv:1006.4106].

\bibitem{11HohenseeChuEtal}
M.A.\ Hohensee, S.\ Chu, A.\ Peters, and H.\ M\"uller,
Phys.\ Rev.\ Lett.\ {\bf 106}, 151102 (2011)
[arXiv:1102.4362].

\bibitem{13Hohensee}
M.A.\ Hohensee, H.\ M\"uller, and R.B.\ Wiringa, 
Phys.\ Rev.\ Lett.\ {\bf 111} 151102 (2013)
[arXiv:1308.2936]. 

\bibitem{12Tasson}
J.D.\ Tasson,
Phys.\ Rev.\ D {\bf 86}, 124021 (2012)
[arXiv:1211.4850].

\bibitem{10Romalis}
J.M.\ Brown, S.J.\ Smullin, T.W.\ Kornack, and M.V.\ Romalis,
Phys.\ Rev.\ Lett.\ {\bf 105}, 151604 (2010)
[arXiv:1006.5425].

\bibitem{10Panjwani}
H.\ Panjwani, L.\ Carbone, and C.C.\ Speake,
in {\it CPT and Lorentz Symmetry V},
edited by V.A.\ Kosteleck\'y, 
World Scientific, Singapore, 2011.

\bibitem{14SchreckMinimal}
M.\ Schreck,
arXiv:1409.1539.

\bibitem{08tables}
V.A.\ Kosteleck\'y and N.\ Russell,
Rev.\ Mod.\ Phys.\  {\bf 83}, 11 (2011)
[2014 edition arXiv:0801.0287v7].

\bibitem{09akjtPRL}
V.A.\ Kosteleck\'y and J.D.\ Tasson,
Phys.\ Rev.\ Lett.\ {\bf 102}, 010402 (2009)
[arXiv:0810.1459].

\bibitem{13akmmNonmin}
V.A.\ Kosteleck\'y and M.\ Mewes,
Phys.\ Rev.\ D {\bf 88}, 096006 (2013)
[arXiv:1308.4973].

\bibitem{causality}
V.A.\ Kosteleck\'y and R.\ Lehnert,
Phys.\ Rev.\ D {\bf 63}, 065008 (2001)
[arXiv:hep-th/0012060].

\bibitem{10classical}
V.A.\ Kosteleck\'y and N.\ Russell,
Phys.\ Lett.\ B {\bf 693}, 443 (2010)
[arXiv:1008.5062].

\bibitem{14Schreck}
M.\ Schreck,
arXiv:1405.5518.

\bibitem{18Finsler}
P.\ Finsler,
{\it \"Uber Kurven und Fl\"achen in allgemeinen R\"aumen},
University of G\"ottingen dissertation, 1918;
Verlag Birkh\"auser, Basel, Switzerland, 1951.

\bibitem{05bcs}
D.\ Bao, S.-S.\ Chern, and Z.\ Shen,
{\it An Introduction to Riemann-Finsler Geometry},
Springer, New York, 2000.

\bibitem{03ShenZermelo}
Z.\ Shen,
Canad.\ J.\ Math.\ {\bf 55}, 112 (2003).

\bibitem{04BRS}
D.\ Bao, C.\ Robles, and Z.\ Shen, 
J.\ Diff.\ Geom. {\bf 66}, 377 (2004).

\bibitem{06Bogoslovsky}
G.Yu.\ Bogoslovsky,
Phys.\ Lett.\ A {\bf 350}, 5 (2006)
[hep-th/0511151].

\bibitem{11pfeifer}
C.\ Pfeifer and M.N.R.\ Wohlfarth,
Phys.\ Rev.\ D {\bf 84}, 044039 (2011)
[arXiv:1104.1079].

\bibitem{11Vacaru}
S.I.\ Vacaru, 
Class.\ Quant.\ Grav. {\bf 28}, 215001 (2011)
[arXiv:1008.4912].

\bibitem{12Laemmerzahl}
C.\ L\"ammerzahl, V.\ Perlick, and W.\ Hasse,
Phys.\ Rev.\ D {\bf 86}, 104042 (2012)
[arXiv:1208.0619].

\bibitem{13mjms}
M.A.\ Javaloyes and M.\ S\'anchez,
Int.\ J.\ Geom.\ Methods Mod.\ Phys.\ {\bf 11} 1460032 (2014)
[arXiv:1311.4770].

\bibitem{14Silva}
J.E.G.\ Silva and C.A.S.\ Almeida,
Phys.\ Lett.\ B {\bf 731}, 74 (2014)
[arXiv:1312.7369].

\bibitem{11Romero}
J.M.\ Romero, O.\ S\'anchez-Santos, and J.D.\ Vergara,
Phys.\ Lett.\ A {\bf 375}, 3817 (2011)
[arXiv:1106.3529].

\bibitem{14Kouretsis}
A.P.\ Kouretsis,
Eur.\ Phys.\ J.\ C {\bf 74}, 2879 (2014) 
[arXiv:1312.4631].

\bibitem{41Randers}
G.\ Randers,
Phys.\ Rev.\ {\bf 59}, 195 (1941).

\bibitem{11akfinsler}
V.A.\ Kosteleck\'y, 
Phys.\ Lett.\ B {\bf 701}, 137 (2011)
[arXiv:1104.5488].

\bibitem{12bipartite}
V.A.\ Kosteleck\'y, N.\ Russell, and R.\ Tso,
Phys. Lett. B {\bf 716}, 470 (2012)
[arXiv:1209.0750].

\bibitem{10Altschul}
B.\ Altschul,
J.\ Phys.\ A {\bf 39}, 13757 (2006)
[hep-th/0602235].

\bibitem{06Lehnert}
R.\ Lehnert,
Phys.\ Rev.\ D {\bf 74}, 125001 (2006)
[hep-th/0609162].

\bibitem{02DCPMD}
D.\ Colladay and P.\ McDonald,
J.\ Math.\ Phys.\  {\bf 43}, 3554 (2002)
[hep-ph/0202066].

\bibitem{12dcpmd} 
D.\ Colladay and P.\ McDonald,
Phys.\ Rev.\ D {\bf 85}, 044042 (2012)
[arXiv:1201.3931].

\bibitem{12fittante}
A.\ Fittante and N.\ Russell,
J.\ Phys.\ G {\bf 39}, 125004 (2012)
[arXiv:1210.2003].

\bibitem{89Beil}
R.G.\ Beil,
Int.\ J.\ Theor.\ Phys.\ {\bf 28}, 659 (1989).

\bibitem{08torsion}
V.A.\ Kosteleck\'y, N.\ Russell, and J.D.\ Tasson,
Phys.\ Rev.\ Lett.\ {\bf 100}, 111102 (2008)
[arXiv:0712.4393].

\bibitem{13NeutronSpin}
R.\ Lehnert, W.M.\ Snow, and H.\ Yan,
Phys.\ Lett.\ B {\bf 730}, 353 (2014)
[arXiv:1311.0467].

\end{thebibliography}
\end{document}